\documentstyle{mn2e}
\input psfig

 \def\be{\begin{equation}}
 \def\ee{\end{equation}}
\def\solmas{{M$_\odot$}}
\def\simless{\mathbin{\lower 3pt\hbox
   {$\rlap{\raise 5pt\hbox{$\char'074$}}\mathchar"7218$}}}   
\def\simgreat{\mathbin{\lower 3pt\hbox
   {$\rlap{\raise 5pt\hbox{$\char'076$}}\mathchar"7218$}}}   
\def\etal{{\rm et al.}}

\def\solmas{{M$_\odot$}}
\def\solm{{M_\odot}}

\def\ms {M_*}
\def\msi {M_{*i}}
\def\tacc {t_{\rm acc}}
\def\vsi {v_{\rm i}}

\def\macc {\dot M_*}

\def\apj{{ApJ}}
\def\apjl{{ApJL}}

\def\mnras{{MNRAS}}

  \makeatletter
  \ifx\CUP@mtlplain@loaded\undefined  
  \makeatother  
    
  \else
    
  \fi

\loadboldmathitalic 
\loadboldgreek      
  
\makeatletter
\ifx\CUP@mtlplain@loaded\undefined  
  \newfont\bit{cmbxti10 at 9pt}
\else
  \newfont\bit{mtbxti10 at 9pt}
\fi
\makeatother

\def\LaTeX{L\kern-.36em\raise.3ex\hbox{a}\kern-.15em
    T\kern-.1667em\lower.7ex\hbox{E}\kern-.125emX}

\newcommand{\gsim}{\mathrel{\hbox{\rlap{\lower.55ex \hbox {$\sim$}}
                   \kern-.3em \raise.4ex \hbox{$>$}}}}
\newcommand{\lsim}{\mathrel{\hbox{\rlap{\lower.55ex \hbox {$\sim$}}
                   \kern-.3em \raise.4ex \hbox{$<$}}}}

\title[Competitive accretion] {Star formation through gravitational collapse {\sl and} competitive accretion}
\author[I. A. Bonnell \etal]
  {Ian A. Bonnell$^1$\thanks{E-mail: iab1@st-and.ac.uk} and  Matthew R. Bate$^2$ \\
$^1$ School of Physics and
  Astronomy, University of St Andrews, North Haugh, St Andrews, Fife,
  KY16 9SS. \\
$^2$ School of Physics, University of Exeter, Stocker Road, Exeter, EX4 4QL \\ }

\date{\today}

\begin{document}

\maketitle

\begin{abstract}
Competitive accretion, a process to explain the origin of the IMF,  occurs when stars in a common gravitational potential accrete
from a distributed gaseous component. Stars located near the centre of the potential
benefit from the gravitational attraction of the full potential and accrete
at much higher rates than do isolated stars. We show that concerns recently raised
on the efficiency of competitive accretion are incorrect as they use globally averaged
properties which are inappropriate for the detailed physics
of a forming stellar cluster. A full treatment requires a  realistic treatment of the cluster potential,
the distribution of turbulent velocities and gas densities. Accreting gas does not travel at the
global virial velocity of the system due to the velocity-sizescale relation inherent in turbulent
gas and due to the  lower velocity dispersion of small-N clusters in which much of the accretion occurs. 
Accretion occurs due to the effect of the local
potential in funneling gas down to the centre. Stars located in the gas-rich centres of such systems 
initially accrete from low relative velocity gas attaining larger masses before needing
to accrete the higher velocity gas. Stars not in the centres of such potentials, or that enter
the cluster later when the velocity dispersion is higher,  do not accrete significantly and thus retain their low-masses.
In competitive accretion, most stars do not continue to accrete significantly such
that their masses are set from the fragmentation process. It is the few stars which
continue to accrete  that become higher-mass stars. Competitive accretion is therefore
likely to be responsible for  the formation of higher-mass stars and can explain the mass distribution, mass segregation and binary frequency of these stars. 
Global kinematics of competitive accretion models  include large-scale mass infall,
with mean
inflow velocities of order $\approx 0.5$ km/s at scales of $0.5$ pc, but 
infall signatures are likely to be confused by  the large tangential velocities and the velocity dispersion present.  Finally,
we discuss potential limitations of
competitive accretion and conclude that competitive accretion is currently the most likely model for the origin of the high-mass end of the IMF.
  
\end{abstract}


\section{Introduction}

One of the primary goals of a complete theory for star formation is to explain the
origin of the distribution of stellar masses, the initial mass function (IMF).  
There
have been many theories for the IMF that have involved a variety of physical
processes from fragmentation (Zinnecker~1984; Larson 1985, Elmegreen 1997, Klessen, Burkert \& Bate 1998; Klessen~2001;Bate, Bonnell \& Bromm 2003; Bonnell, Clarke \& Bate 2006), turbulence (Elmegreen~1993;  Padoan \& Nordlund~2002) accretion (Zinnecker~1982, Larson~1992, Bonnell \etal~1997,2001b; Klessen \& Burkert 2000; Bate \& Bonnell~2005), feedback (Silk~1995, Adams \& Fatuzzo 1996),
magnetic fields (Shu. Li \& Allen 2004) and a
combination of these (Adams \& Fatuzzo 1996).  The simplest possible mechanism 
that can explain the origin of the IMF is one that relies primarily on the physics of gravity,
namely fragmentation, accretion and dynamical interactions.  (Bonnell, Larson \& Zinnecker 2006). In this scenario, gravitational fragmentation of molecular clouds sets the mean stellar mass
at $\approx 0.5 \solm$. Lower-mass stars and brown dwarfs arise due to the small Jeans masses produced
in collapsing regions (filaments, discs), followed by stellar interactions and potentially
ejections to 
stop any subsequent accretion (Bate, Bonnell \& Bromm 2002a; Reipurth \& Clarke 2001). Higher-mass stars form due to  continued accretion in a clustered
environment where the overall system potential funnels gas down to the centre of the potential, to be accreted by the proto-massive stars located there. Not only can this
reproduce the stellar IMF, but it also is able to account for the mass segregation of young stellar clusters
and the binary properties of low and high-mass stars (Bonnell \& Bate 2005; Bate, Bonnell \& Bromm 2002b).

Turbulent fragmentation has also been suggested as an alternative mechanism whereby the masses
of stars are determined by the size and densities of turbulent shocks (Padoan \& Nordlund 2002).
Lower velocity  shocks produce weak but large shocks and hence result in higher mass clumps/stars
whereas high-velocity shocks produce strong but narrow shocks and hence lower mass clumps/stars.
One difficulty with this model (see also Ballesteros-Paredes \etal 2006) is that the massive stars
should form well separated and hence in relative isolation, not in the centre of dense stellar clusters.
Indeed, in turbulent fragmentation, it is the lower-mass cores that are expected to form in such
densely packed regions.

Competitive accretion relies on  the inefficiency of fragmentation such that there is a large common reservoir
of gas from which the protostars can accrete (e.g. Bonnell \etal 2001a). 
Observations of pre-stellar structures and of young stellar
clusters both support this view with the large majority of the total mass being in a distributed gaseous form (Motte \etal~1998; Johnstone \etal~2000, 2004; Lada \& Lada 2003; Bastian \& Goodwin 2006).
The second requirement is that the gas be free to move under the same gravitational acceleration as the stars.
If the gas is fixed in place due to magnetic fields then accretion will be limited.
When these two requirements are filled, the dynamical timescale for accretion and evolution are similar such that a significant
amount of gas can be accreted. Models of the formation of a stellar cluster show that the initial fragmentation
produces objects of order the Jeans mass of the cloud, even in the presence of turbulent velocities (Bonnell, Vine \& Bate
2004). Accretion of nearby low relative velocity material within their tidal radii increase these masses somewhat and helps
produce the shallow IMF for low-mass stars (Bonnell \etal 2001b; Klessen \& Burkert 2000). The stars then fall together to
form small stellar clusters which grow by accreting stars and gas (Bonnell \etal 2003, 2004).
This produces a strong correlation between the number of stars in the cluster and the mass of the most massive star it contains, as is found in young stellar clusters (Weidner \& Kroupa 2006). Stars near the centre of the clusters benefit from the full potential which funnels
gas down to them, increasing the local gas density.
Once the stars establish a stellar dominated region in the centre of the clusters, they
virialise and have higher relative gas velocities such that the Bondi-Hoyle radius, based on local parameters, now determines the accretion rate. Accretion in this regime produces the higher-mass stars (Bonnell \etal 2004) 
as well as the Salpeter-like high-mass IMF (Bonnell \etal 2001b; Bonnell \etal 2003). Furthermore, the stellar dynamics
maintain the accreting massive star in the centre of the system, and hence with a low
relative velocity compared to the cluster potential.

Recently, Krumholz, Klein \& McKee (2005a) have cast some doubt on this process by claiming that,
accretion in such environments cannot significantly increase a star's mass and therefore does not
play a role in establishing the stellar IMF.  In part this is correct as with competitive accretion,
most stars do not continue to accrete. It is the few that do that are important in terms of forming higher
mass stars and the IMF.
We have  reanalysed our results in view of their
work in order to show why accretion does occur to form higher-mass stars in our simulations and to establish if this
is a realistic outcome of star formaiton. The principle difference is that the Krumholz \etal\ 
analysis 
used global parameters for the gas properties which can be significantly different from the local
properties of a forming stellar cluster. We find that using global properties significantly underestimates the correct accretion rates for the few stars that achieve higher masses.  In \S 2 we discuss the Krumholz \etal\ analysis and point out some
limitations in their approach. In \S 3 we reanalyse our numerical simulations. In \S 4 we discuss
observational constraints of the models and in \S 5 we investigate potential limitations of
competitive accretion. Finally, we summarize our current understanding of competitive accretion in \S 6.

\section{Analytical estimates of the accretion rates}

The Krumholz \etal~(2005a) approach considers accretion in a molecular cloud
supported against self-gravity by its internal turbulent motions.
In order to make the problem analytically tractable, they use globally averaged
properties of the gas and relative velocities. In order to do this, they first assume that the
gas is not globally self-gravitating.  This implies that 
that the gas mass is then less than the mass in stars, and under such circumstances
it is obvious that accretion cannot significantly alter a stars mass. They also neglect the larger scale cluster potential as being anything
but a boundary to keep the gas within the system. 
Together, these assumptions ignore the similar accelerations that both the gas and star undergo in a
cluster potential, and which to a large degree determine the relative velocities.

The accretion rates can  be estimated from the Bondi-Hoyle formalism where
the accretion rate, $\macc$ is given by
\be
\label{eqBH}
\macc \approx 4 \pi \rho \frac{\left( G \ms \right )^2}{v^3},
\ee
where $\ms$ is the stellar  mass, $\rho$ the gas density, $G$ the gravitational constant and $v$ the
relative velocity of the gas.  Using such a formalism thus depends on having
accurate values for the gas density, stellar masses and relative velocities. Unfortunately,
Krumholz~\etal\~(2005a) use globally averaged estimates
for the gas density and the velocity instead of the local variables as used in the
Bonnell \etal (2001a, 2004) numerical simulations. This causes the
many order of magnitude difference in the calculated accretion rates.

A stellar cluster or other self-gravitating system is not of uniform density but has
a significantly increased density near the centre of the gravitational potential. For example,
the mass density in the centre of the Orion nebula Cluster (ONC) is approximately 50
times the mean mass density of the system (Hillenbrand \& Hartmann 1998). Similar
central condensations are produced in simulations of turbulent fragmentation and cluster formation (Bonnell \etal~2003). Indeed,
in competing models of massive star formation through the pressure-induced compression
of turbulent cores,
the central mass density is much higher than this (McKee \& Tan 2003). Using a global
virial velocity for the relative gas velocity is also problematic as it neglects that turbulence
has a velocity size-scale relation and thus much local gas travels at significantly lower
velocities. Furthermore, if a core is formed through turbulent shocks, then much of the
kinetic energy in the surrounding gas will also have been dissipated. The velocities
of the forming stars are also not large for the same reason and even once they
fall into local potential minima to form small-N clusters, the velocity dispersion in these systems
is low. A forming cluster of 10 stars in a 0.1 pc radius will have a velocity dispersion of order
0.4 km/s instead of 2 km/s for the ONC. In terms of initial masses, Krumholz \etal\ use
$\ms=0.1\solm$ rather than in the simulations where the mean initial masses
are $\ms\approx 0.5\solm$. Thus, in terms of a Bondi-Hoyle accretion yields
\be
{\macc\over \dot M_{*, \rm glob}} = {\rho \over \rho_{\rm glob}} \left({\ms \over M_{*, \rm glob}}
\right)^2 
\left({v \over v_{\rm glob}}\right)^{-3},
\ee
which when we incorporate the differences between local and global values gives
\be
{\macc\over \dot M_{*, \rm glob}} = 50 \times 5^2 \times 5^3 \approx 1.5 \times 10^5.
\ee
This difference of $1\times 10^5$ in the accretion rates is more than sufficient to explain why the Krumholz \etal analytical analysis significantly underestimated the accretion rates from the numerical simulations. Thus, for example, using a relative gas velocity of $v=0.4$ km/s, a stellar mass of
$\ms=0.5\solm$ and a gas density of $\rho=10^{-17}$ g cm$^{-3}$ ($\approx$ the mass density in the core of the ONC), the accretion rate is then $\macc \approx 10^{-4}
\solm$ yr$^{-1}$; whereas with $v=2$ km/s, 
$\ms=0.1\solm$ and $\rho=2 \times10^{-19}$ g cm$^{-3}$, in line with Krumholz \etal (2005a),
the estimated accretion rate is then  $\macc \approx 10^{-9}
\solm$ yr$^{-1}$. These values are summarised in Table~1. Such a large difference in accretion rates for seemingly small changes in the
physical parameters highlights the danger of using global properties to predict what is, after all, a very local physical process. It also highlights the power of competitive accretion as small
changes in the local properties such as an acceleration from a stellar encounter
and even the ejection into a lower gas-density region, can effectively halt the continued accretion.

\begin{table} \caption{\label{simulations} Accretion rates for global and local properties}
\begin{tabular}{c|c|c}
\hline
Property & Global & Local \\
Gas density (g cm$^{-3}$) & $2 \times 10^{-19} $ & $10^{-17}$ \\
Velocity dispersion (km s$^{-1}$) & 2 & 0.4 \\
Stellar mass (\solmas) & 0.1 & 0.5 \\
Accretion rate (\solmas\ yr$^{-1}$) & $10^{-9}$ & $10^{-4}$ \\
\hline
\end{tabular} \end{table}

\begin{figure}
\centerline{\psfig{figure=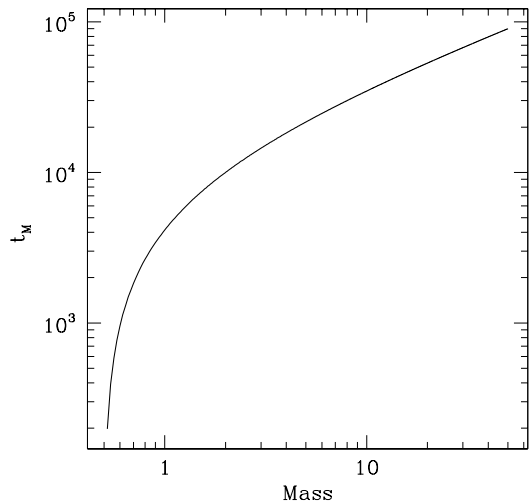,width=8.truecm,height=8.truecm}}
\caption{\label{taccvsm}  The accretion timescale in years, from equation (\ref{tacceq}), is plotted as a function of the final stellar mass in \solmas. The plot assumes an initial mass of $0.5\solm$ and a gas density of $\rho=10^{-17}$ g cm${-3}$. The accretion timescale is directly proportional to the inverse of the stellar density.}
\end{figure}

\begin{figure*}
\centerline{\psfig{figure=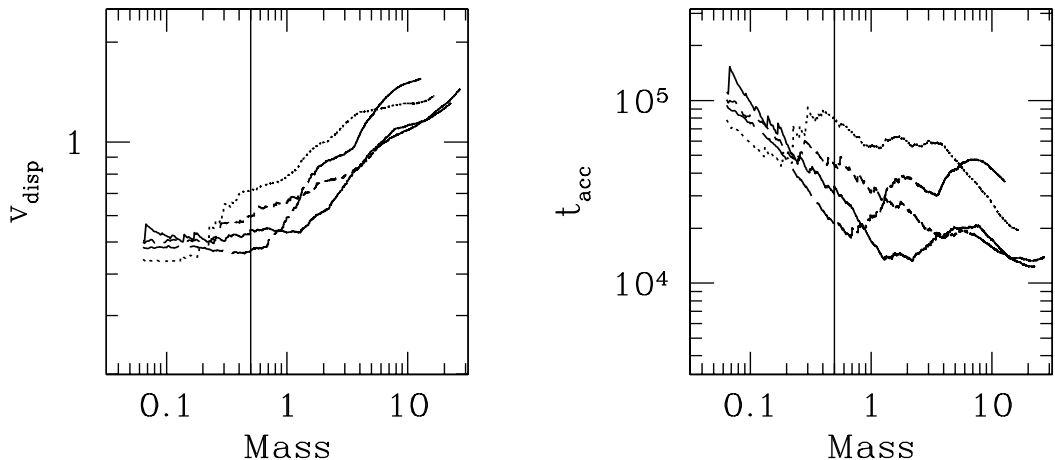,width=16.truecm,height=15.truecm,rheight=7.5truecm}}
\caption{\label{vdisp}  The 3-D velocity dispersion of the accreted gas is plotted (left) as a function of the cumulative accreted gas mass (in \solmas) for four massive stars. The velocity dispersion is low for low masses
due to the nature of turbulence and the low velocity dispersions in small-N clusters where much
of the accretion occurs. The velocity dispersion increases with accreted mass due to the
velocity-sizescale relation for turbulent clouds. The accretion timescale, $\tacc\approx \ms/\macc$, 
is plotted as a function of accreted mass (right) showing that accretion occurs on relatively short timescales and thus determines the final masses of high-mass stars.}
\end{figure*}

We can estimate the timescale for accretion to form a massive star by considering that a
star forms with masses typical of the local Jeans mass which is near the median stellar mass ($\ms\approx 0.5\solm$). We further consider that the relative gas velocity is initially low as expressed
above (0.4 km/s) and that it follows a turbulent velocity sizescale relation of the form (Larson 1981; Heyer \& Brunt 2004)
\be
v \propto R^{1/2}.
\ee
Assuming the mass distribution follows that of a centrally condensed $\rho \propto R^{-2}$ sphere
such that $M\propto R$ (see Fig.~4 of Bonnell \etal 2004),  provides a relation between the turbulent velocity
and the mass of the region considered
\be
v \propto M^{1/2}.
\ee
Combining this with equation (\ref{eqBH}) above gives
\be
\frac{d \ms}{dt} \approx 4 \pi \rho \frac{\left( G \ms \right)^2}{{(\vsi)}^3 \left(\frac{\ms}{\msi}\right)^{3/2}},
\ee
which can be integrated to yield an accretion timescale to form a star of final mass $\ms$ of
\be
t_{\rm M}= \frac{2\vsi^3\left( \ms^{1/2} - \msi^{1/2}\right)}{4 \pi \rho \msi^{3/2} G^2}.
\ee
For  typical values found in the numerical simulations of $\vsi\approx 0.5$ km/s, $\rho_{\rm clus}\approx 10^{-17}$ g cm$^{-3}$ and $\msi\approx 0.5 \solm$, this reduces to
\be
\label{tacceq}
t_{\rm M} \approx 10^4 \left[\left(\frac{\ms}{0.5 \solm}\right)^{1/2} -1\right] {\rm years}.
\ee
This accretion timescale is plotted as a function of the final stellar mass in Figure~\ref{taccvsm}.
Thus to form a $10\solm$ star requires $\approx 3.3\times 10^4$ years while a $50\solm$ star
requires $\approx 9\times 10^4$ years. These are certainly reasonable numbers suggesting that
competitive accretion can indeed account for the formation of higher mass stars.
There is still the issue of radiation pressure impeding the formation of stars greater than
10-40 \solmas (Yorke \& Kr\"ugel 1979; Wolfire \& Casinelli 1986), but sufficient current hypotheses exist to overcome this 
obstacle, including disc
accretion and radiation beaming (Yorke \& Sonnhalter 2002), Rayleigh-Taylor instabilities
in the accretion flow (Krumholz \etal 2005b) and stellar collisions (Bonnell, Bate \& Zinnecker 1998; Bonnell \& Bate 2002) and binary mergers (Bonnell \& Bate 2005).

\section{Numerical models of competitive accretion}

One of the advantages of using the smoothed particle hydrodynamic (SPH) technique 
to follow star formation is that it is Lagrangian and therefore allows for the direct tracing of the
fluid flow. We can thus directly probe the physical conditions of the accretion and assess
what is occurring. 
In this section, we re-analyse the results from our previous simulation of the fragmentation
of a turbulent cloud and the formation of a stellar cluster (Bonnell \etal~2003). It is this
simulation which showed how massive stars can form through competitive accretion 
in the centre of the forming clusters (Bonnell \etal~2004). 

The SPH technique uses sink-particles (Bate \etal~1995) to follow the formation of stars
and any continuing gas accretion onto them by adding the mass (and linear/angular momentum) of accreted particles to that of the sink-particle. It is therefore straightforward to track the evolution of the gas
before it was accreted. In this way, we can determine the physical properties of the gas that
formed a given star at all stages before it is accreted. For example, the physical distribution of this
mass is shown in Bonnell \etal~(2004). Here, we  analyse the velocity dispersion of the gas
in order to determine how the accretion proceeds.

In Figure~\ref{vdisp} we plot the 3-D velocity dispersion, at the time the sink-particle forms, of the gas which eventually comprises four massive stars. This shows the velocity dispersion of all the gas which is yet to accrete
onto the star at the time the star first forms, and is plotted as a function of the mass in order of its accretion history. This therefore neglects any subsequent decrease in the
gas's kinetic energy due to shocks. We  see from figure~\ref{vdisp}  that the velocity
dispersion is flat at $\approx 0.5$ km/s for low-masses until $\ms\approx 0.5\solm$. This is 
approximately the mean fragment mass ($0.5 \solm$)
in the turbulent simulation and the velocity dispersion can be understood as being due to the gravitational collapse
of the fragment at the sizescale of the fragment of several hundred AU. Accretion and higher-mass star formation accompanies
the formation of a stellar cluster (Bonnell \etal\ 2004), such that the growth from low-masses
starts in small-N clusters where the velocity dispersion is intrinsically low. 
Beyond this mass, the velocity
dispersion increases with mass, as one expects for a turbulent medium, until nearly reaching the system velocity dispersion of 2 km/s. 
The actual increase in the velocity dispersion approximately follows the $v\propto M^{1/2}$,
as discussed above. This is in keeping with a turbulent medium where the
gas kinematics have not been overly affected by any dissipation, and further reenforces
our choice of the initial velocity dispersion above of $0.5$ km/s.

We can use the above velocity dispersion as a function of accreted gas mass to 
evaluate typical accretion timescales. Figure~\ref{vdisp} also shows the accretion
timescale for a star to accrete its own mass from
\be
\tacc \approx \frac{\ms}{\macc},
\ee
where $\macc$ is calculated from equation~(\ref{eqBH}) using the velocity dispersion and
accreted mass from the simulations and assuming a gas density of $10^{-17}$ g cm${-3}$.
At any given time during the simulation after star formation has commenced, at least $50-100 \solm$ of gas has densities greater than this value.
We can see from Figure~\ref{vdisp} that the accretion timescale is between $10^4$ to $10^5$
years for masses greater than the initial fragment mass of $\approx 0.5 \solm$.   From
equation~(\ref{tacceq}),  the estimated timescale for a star to double its mass is
$\tacc \approx 1.4 \times 10^4$ years, on par with some of the smaller values for $\tacc$ in 
Figure~\ref{vdisp}. These timescale
are sufficiently short that there is no difficulty forming higher mass stars from continued accretion in a clustered environment. The simulation formed two $\approx 30 \solm$ stars and a number of $m>10\solm$ stars in a few $\times 10^5$ years.
The accretion
rates ($\propto M^2/v^3$) actually increase with stellar mass, even though the gas now comes with larger
velocity dispersions. This is due to the $v\propto M^{1/2}$ seen in Figure~\ref{vdisp}.
We can therefore conclude that it is essential to use local quantities to correctly evaluate the accretion rates.

\section{Infall and large-scale kinematics}

Krumholz \etal (2005a) have argued that one way that competitive accretion could occur is
if the system was significantly subvirial such that the relative gas velocities are then very small,
and the whole system is in a state of radial collapse.
We have seen above that this is not necessary due to the velocity-sizescale relation observed
in turbulent molecular clouds such that smaller scales naturally have lower velocity dispersions.
Furthermore, as the initial burst of accretion occurs in small-N clusters where the stellar
velocity dispersion is low, accreting stars can grow to higher-masses before needing to accrete
higher velocity dispersion gas. Nevertheless. there must be some inflow into the central regions
in order for the mass to arrive at where the massive star is forming. This, in fact is not unique
to competitive accretion models as any model of massive star formation in the centre of a stellar
cluster must at some point have inflow of mass to the cluster centre. In the McKee \& Tan (2003)
model for example, this inflow is simply presumed to have occurred at an earlier stage.

We have analysed the gas kinematics from a simulation of a forming stellar cluster (Bonnell \etal 2003). We find, in contradiction to the prediction of Krumholz \etal (2005a), that the system 
maintains approximate global virial equilibrium (ie, $\alpha_{\rm vir} = E_{\rm kin} / |E_{\rm pot}| \approx 0.5$).
This can be easily
understood in the terms of gravitational dynamics where even a cold collisionless system
forms a virialised system at half its initial radius. In terms of a forming stellar cluster, as long
as the gas is highly structured, and especially as it contains significant initial tangential motions, then
it too will maintain a near-virial configuration. 
The kinetic energy of the system is initially equal to the potential energy. The subsequent 
decay of kinetic energy in shocks allows the system to become globally bound. Once gravity
forms local potential wells into which the stars and gas fall, the gravitational acceleration
increases both the radial and the tangential velocities such that the gas has larger, and still
highly disordered motions. In order to attain the evolution suggested in Krumholz \etal (2005a),
there would have to be no initial tangential motions or no conservation of angular momentum during infall. 

\begin{figure}
\centerline{\psfig{figure=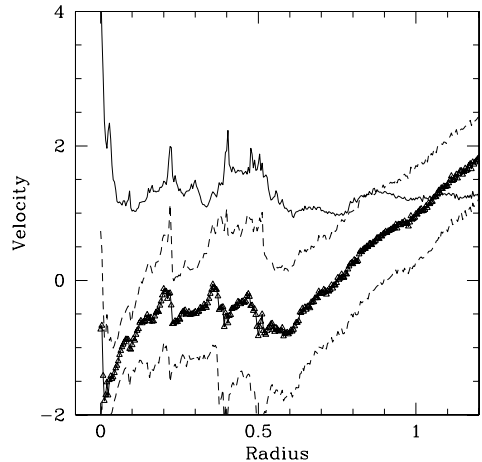,width=8.truecm,height=8.truecm}}
\caption{\label{gaskin}  The mean radial velocity (km/s) in shells is plotted as a function of the
distance (in pc) from the forming massive star (symbols). The dashed lines denote the velocity
dispersion around the mean radial velocity while the solid line is the mean tangential
velocity. Note that the inward radial velocity is smaller than both the tangential velocity and 
the dispersion in the radial velocity.}
\end{figure}

In order to  quantify the kinematics, we plot 
in Figure~\ref{gaskin}  the mean radial and tangential velocities as a function
of radius from the accreting massive star. The SPH particles are binned as a function of
radial shells and the mean radial and tangential velocities are calculated along with the
dispersion in radial velocity. The mean radial velocity is  inwards from $\approx 0.6$ pc 
with a typical value of $0.5-1$ km/s. The velocities outside this radius are generally outwards
due to the large kinetic energy in the initial conditions. The dispersion in the radial velocity is actually larger ($\pm 1$ km/s)
than this inward velocity indicating that the motions are very chaotic. 
The tangential
velocities are also larger than the radial velocity demonstrating that the system is not undergoing a
simple collapse process but involves significant chaotic/turbulent motions and is kinematically hot.
The system is significantly different from the cold collapse model suggested by Krumholz \etal(2005a).

Observational studies have searched for infall signatures in massive star formation. Recently, Fuller \etal (2005), Motte \etal\ (2005)  and Perreto \etal (2006) have  detected of such motions through asymmetric line profiles,
and by spatially resolved kinematics, respectively. The estimated infall velocities are of the order
0.1 to 2 km/s, in rough agreement with the mean inward velocities seen in the numerical simulations. 

\begin{figure}
\centerline{\psfig{figure=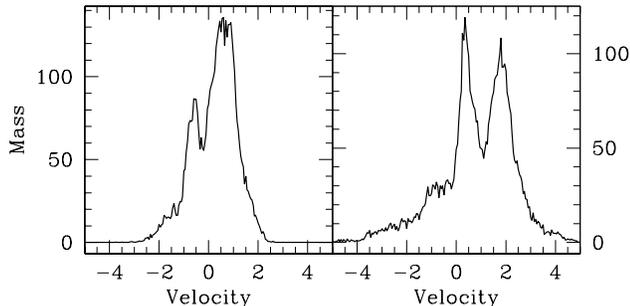,width=9.truecm,height=9.truecm,rheight=4.5truecm}}
\caption{\label{linediag}  The mass of the gas within a projected radius of 0.1 pc from the forming massive star
is plotted as a function of its line-of-sight velocity at two different times during the evolution
when the most massive star has 4 and 12 \solmas, respectively. The mass is plotted in units
of \solmas pc$^{-2} ({\rm km/s})^{-1}$ while the velocities are in units of km/s.
Although local velocity dispersions are small ($\approx 0.5$ km/s Figure~\ref{vdisp}), the line of sight velocity
dispersions are significantly larger ( few km/s ) due to the multiple clumps in the line of sight. 
}
\end{figure}

Observed linewidths in regions of massive star formation have been  found to be of order a few km/s
(e.g. Garay 2005; Beuther \etal 2006), This is significantly
larger than the values discussed above  in terms of the local velocity dispersion of the gas
on relatively small scales. The difference can be understood as being due to the large
lengthscale of the core probed along the line of sight. Even when small areas of the core are observed, the gas is likely to lie along an extended. This effect shown in  Figure~\ref{linediag} where we plot the line of sight velocity distribution of the gas mass within a projected areas of radius $0.1$ pc 
centred on the
forming massive star. This effective velocity profile for an optically thin tracer has a line width
of several km/s in agreement with observations. Thus, although the small-scale velocity dispersion
in a 3-D volume can be fairly low, the line of sight velocity dispersion in a 2-D projected area
is much larger due to the greater sizescale probed. The existence of multiple distinct clumps
can also be seen in the right hand panel where two clumps are now kinematically distinct.
Such highly structured profiles which can be observed along some lines of sight are
due to the hierarchical fragmentation of the turbulent cloud (Bonnell \etal\ 2003).


\section{Potential limitations of competitive accretion}
We have seen above that competitive accretion arises naturally from the fragmentation of a turbulent
molecular cloud where there is a common (cluster) potential such that stars near the centre of the potential accrete from higher density gas and generally have lower relative velocities. The
relative gas velocities are low due to the nature of turbulence and its velocity-sizescale relation
with lower velocities on smaller sizescales. In order for competitive accretion not to work as
suggested by Krumholz \etal (2005a), requires that the local gas velocities are very high, generally
as high as the turbulent velocity on the largest scale of the cloud. Neglecting that this would
violate the velocity-sizescale relation ($v \propto R^{1/2}$), we discuss here potential mechanisms
that could decouple the gas and stellar kinematics.

In order to completely decouple the stellar and gas kinematics requires a driving source
for the gas motions. Such a driving source, potentially related to maintaining the turbulent support 
for many dynamical lifetimes, is often speculated to  be the internal feedback from low-mass stars. The problem  is that although
the energetics from low-mass stellar jets and outflows are sufficient to offset the decay
of turbulent energy, it is very unclear if they
can input sufficient kinetic energy into the system without completely removing the gas (Arce \etal 2006).
Jets are commonly found to escape the bounds of their natal clouds suggesting that the
majority of their momenta is deposited at large distances, outside the main star forming region
Stanke \etal 2000; see also Arce \etal 2006; Bally \etal 2006). 
Even the intrinsically isotropic feedback from high-mass stars escapes in preferential directions due to the non-uniform gas distributions (Dale \etal~2005; Krumholz \etal~2005c). The feedback
decreases the accretion rates but does not halt accretion. Indeed, in the toy-model feedback simulation
 by Li \& Nakamura (2005), the energy injection is sufficient to
disrupt the core but is quickly damped from the system allowing it to continue its unimpeded collapse.

A further problem is one of balancing the timescales. 
Energy injection from jets and outflows is relatively quick with typical  dynamical times of the flows 
is generally of order 
$10^3$ to $10^4$ years (e.g. Reipurth \& Bally 2001; Beuther \etal\ 2002).  Giant outflows have longer dynamical timescales but considering outflow velocities of 10 to 100 km/s,
the energy injection timescale of an outlfow before it leaves he limited size of
a forming stellar cluster  
is at most a few $10^4 years$. This timescale is  smaller than the dynamical timescale
of the system as a whole (few $10^5$ years). It is difficult
to envision how the feedback can be exactly tuned to support the system
without disrupting it. If excessive energy is deposited by the feedback, there is
no opportunity for the cluster to adapt before the gas is dispersed. Likewise, if
insufficient energy is deposited, then the system will continue to form stars.
As the timescale for star formation is similar to the dynamical timescale of the cluster,
it cannot be halted on the dynamical timescales of the outflows such that feedback is highly
likely to overshoot the energy balance and completely unbind the system. Indeed,
observations suggest that this is common (Arce \etal 2006) and that there is actually
no need for long lifetimes as star formation appears to generally occur on dynamical
timescales (Elmegreen 2000).

\section{Conclusions}

We have shown that concerns raised about the efficacy of competitive accretion are misplaced
due to the misuse of global variables to estimate what is a local physical process.
In competitive accretion, it is the few stars located early on in the centre of local stellar
systems that become higher-mass stars. Competitive accretion starts with the accretion of low relative velocity gas 
due to the nature of turbulence and the low velocity dispersion in small-N clusters. Accretion
proceeds to higher relative velocity gas once the star has attained sufficient mass
to maintain a high accretion rate. Stars that enter a stellar cluster later or that reside
far from the centre of the potential do not accrete significantly and hence lose the
competition to accrete from the distributed gas. In this way, competitive accretion sets
the distribution of stellar masses by determining which stars accrete to attain higher-masses
and which do not so that they remain lower-mass stars. Although infall does
occur, as it needs to in any model for massive star formation in the centre of a cluster, 
emission line signatures are likely to be confused due to the large chaotic and tangential
motions present. The forming cluster never approaches the status of a cold, radially collapsing system.

\section*{Acknowledgments}
 We thank Henrik Beuther, Cathie Clarke, Ralf Klessen, Richard Larson, Jim Pringle and Hans Zinnecker for discussions and comments
 which helped formulate this letter. We acknowledge the contribution of the  U.K. Astrophysical Fluids Facility (UKAFF)
 in providing a facility to test and refine our models. MRB is grateful for the support of a Philip Leverhulme Prize.

\end{document}